\documentclass{aa}
\usepackage{natbib,graphicx,lscape}
\def\aj{\rm AJ}
\def\araa{\rm ARA\&A}
\def\apj{\rm ApJ}
\def\apjl{\rm ApJ}
\def\apjs{\rm ApJS}

\def\aap{\rm A\&A}
\def\aapr{\rm A\&A~Rev.}
\def\aaps{\rm A\&AS}
\def\mnras{\rm MNRAS}

\def\pasj{\rm PASJ}

\newcommand{\nh}{\hbox{$N_{\rm H}$}}

\newcommand{\n}[1]{$N_\mathrm{#1}$}
  
\begin{document}
 
\title{ROSAT PSPC view of the hot interstellar medium of the Magellanic Clouds}

\titlerunning{Hot interstellar medium of the MCs}

\author{M.\ Sasaki \and F.\ Haberl \and W.\ Pietsch}
 
\authorrunning{Sasaki et al.}                                                  
 
\offprints{M. Sasaki, \email{manami@mpe.mpg.de}}
 
\institute{Max-Planck-Institut f\"ur extraterrestrische Physik,
               Giessenbachstra{\ss}e, Postfach 1312, 85741 Garching, Germany}
 
\date{Received March 15, 2002; accepted June 17, 2002}
 
\abstract{
Diffuse X-ray emission from the Magellanic Clouds (MCs) is
studied by using all the archival data of pointed ROSAT Position
Sensitive Proportional Counter (PSPC) observations. 
For this purpose, contributions from the point and
point-like sources in the ROSAT High Resolution Imager (HRI) and 
PSPC source catalogues
are eliminated and periods of high solar activity are excluded.
The spectral analysis yielded characteristic temperatures of 
$10^{6} - 10^{7}$~K for the hot thin plasma of the ISM which 
extends over the whole Large Magellanic Cloud (LMC) and the Small
Magellanic Cloud (SMC). The total unabsorbed luminosity in the 
0.1 -- 2.4~keV band within the observed area amounts to
$3.2 \times 10^{38}$\,erg\,s$^{-1}$ in the LMC and
$1.1 \times 10^{37}$\,erg\,s$^{-1}$ in the SMC, each 
with an uncertainty of $\sim-40$\%, $+100$\%.
The X-ray luminosity of the LMC is comparable to that of 
other nearby galaxies with pronounced star formation.
In the LMC, hot regions were found especially around the
supergiant shell (SGS) LMC\,4 and in the field covering SGS LMC\,2 and LMC\,3.
Highest temperatures for the SMC were derived in the southwestern part of the 
galaxy.
The diffuse X-ray emission is most likely a superposition of the emission from
the hot gas in the interior of shells and supershells as well as
from the halo of these galaxies.
}
 
\maketitle
 
\section{Introduction}                                                       

The small distances to the Magellanic Clouds (MCs) allow us to separate 
the emission from distinct sources  
from that arising from surrounding gas within the galaxies.
Not only the discrete point sources
and extended supernova remnants (SNRs), 
but also diffuse emission coming from the interstellar
medium (ISM) can be observed and studied in detail. In the interstellar space, 
stars are born out of the densest regions, whereas massive stars transfer 
matter back to the ISM in stellar winds and supernova explosions. Therefore,
the diffuse component of the X-ray emission from galaxies will give us clues 
for a better understanding of the interaction between stars and the ISM
as well as for the matter cycle within the galaxies. 

With Einstein and ROSAT, high resolution X-ray imaging became possible and
revealed diffuse X-ray emission in the Galaxy, the MCs, and other nearby 
galaxies \citep[e.g.][]{1984ApJ...286..491F,1984ApJ...286..144W,1991Sci...252.1529S,1993namc.meet...59P}. 
It indicated the existence of a very hot component in the interstellar
medium (ISM) with temperatures of $\sim10^{6}$~K besides
the cold gas observed in radio as well as warm components seen in the
optical or UV. First supergiant shells (SGS) of relatively cold, ionized matter
in the MCs were identified on H$\alpha+$[\ion{N}{ii}] images as filamentary 
structures \citep{1978A&A....68..189G,1980MNRAS.192..365M}. 
Shells are interpreted to result from matter swept up by expanding 
gas, and can be also seen in \ion{H}{i} emission maps as dense regions around voids in 
the neutral hydrogen distribution. 

Based on Einstein IPC data, very hot gas in the Large Magellanic
Cloud (LMC) was mapped by
\citet{1991ApJ...374..475W}, and further detailed analysis of the diffuse X-ray
emission was performed \citep[e.g.][]{1991ApJ...373..497W,1991ApJ...379..327W}.
SGS LMC\,2 \citep{1980MNRAS.192..365M} is one of the
supergiant shells discovered in the LMC and is located next to the star
formation region 30 Doradus.
\citet{1991ApJ...379..327W} analyzed the Einstein IPC data 
of the SGS LMC\,2 comparing it to infrared and \ion{H}{i} observations,
and found a ring of X-ray emission with a temperature of about 
$5 \times 10^{6}$~K 
within a cavity in the \ion{H}{i} map.

\citet{1994A&A...283L..21B} analyzed ROSAT Position Sensitive
Proportional Counter (PSPC)
pointings covering the
northern part of SGS LMC\,4, which is located in the north
of the LMC. They derived a temperature of $2.4 \times 10^{6}$~K by fitting a
thermal plasma spectrum. As for SGS LMC\,2,
ROSAT and ASCA observations made it possible to measure a plasma temperature of $k T 
\approx 0.1 - 0.7$~keV \citep{2000ApJ...545..827P}. 
Merged ROSAT PSPC images of the MCs were presented by
\citet{1994ApJ...436L.123S} and \citet{1999IAUS..190...32S}. For the LMC,
140 PSPC pointings were used and 20 for the Small Magellanic Cloud (SMC), 
and 
images were created in different spectral bands of the detector.
In the LMC, a hot plasma was found along the optical bar with temperatures
between $4 \times 10^{6}$ and $8 \times 10^{6}$~K increasing from west to
east. In the SMC no pronounced diffuse X-ray emission was detected.

Both the LMC and the SMC were observed by ROSAT \citep{1982AdSpR...2..241T} 
in a period of over eight years in nearly 900 pointings, which covered 
the MCs almost completely.
Therefore, a thorough study of the X-ray emission from the MCs
has been started. The aim of this work was to establish a detailed picture of  
the high energy processes within a galaxy by producing a catalogue of
sources in the MCs in the ROSAT band 
\citep[see][]{1999A&AS..139..277H,2000A&AS..142...41H,2000A&AS..143..391S,2000A&AS..147...75S}
and by analyzing the hot component of the interstellar medium.
Since objects which can be detected in X-rays at the distances of the MCs are
SNRs or binary systems including objects at the final 
stages of stellar evolution (supersoft sources, SSSs and X-ray binaries, XRBs),
their distribution in combination with the structure and physical
state of the ISM will indicate the region within the galaxies
developing most actively and help us to understand the evolutionary history of
the MCs. This work presents the spectral analysis of the complete
ROSAT PSPC data 
unveiling the structure of the diffuse X-ray emission of the MCs with a 
spatial resolution of 15\arcmin\ $\times$ 15\arcmin\,.

\section{Data}

The archival data of the ROSAT PSPC are best suited for spectral analysis of 
the MCs as a whole, because there were pointed observations 
with an almost full coverage and high photon 
statistics. In total 223 pointings for the LMC in a field of 10\degr\ $\times$
10\degr\ around RA = 05$^{\rm h}$ 25$^{\rm m}$ 00$^{\rm s}$, Dec =
--67\degr\ 43\arcmin\ 20\arcsec\ (J2000.0) and 31 for the SMC in a field of 
5\degr\ $\times$ 5\degr\ around
RA = 01$^{\rm h}$ 00$^{\rm m}$ 00$^{\rm s}$, Dec = $-73\degr\ 00\arcmin\
00\arcsec$ (J2000.0) with 
single pointing exposure times of up to 
65\,ks 
were analyzed using the EXSAS software package 
\citep{1994exsas.....Z}.

Since the diffuse X-ray emission has a very low surface brightness and is 
extended over the field of view, it is important to eliminate the emission 
from point sources as well as the background emission very carefully.
The background has to be modeled and can not be directly taken from the
same observation as is common in point source analysis. 
In the course of spectral analysis, the background modeling
is also one of the major topics,
as will be shown in Sect.\,\ref{bgfgspec}. But here, 
we first discuss the background components which can be eliminated 
during data preparation. 

\subsection{Discrete Sources}

Based on ROSAT PSPC and High Resolution Imager (HRI) catalogues of distinct X-ray 
sources in the MCs
\citep{1999A&AS..139..277H,2000A&AS..142...41H,2000A&AS..143..391S,2000A&AS..147...75S}, 
the contamination from all point and point-like X-ray sources in the ROSAT band 
was eliminated from the data. The extraction radii were computed 
according to the point spread function of the observations on the one hand, 
and the brightness and the extent of the sources on the other hand. 

The ROSAT catalogues of discrete sources are complete to luminosities of 
$\sim3 \times 10^{33}$~erg~s$^{-1}$. 
Very faint X-ray sources like 
cataclysmic variables or RS CVn stars with luminosities lower than
$\sim 10^{33}$~erg~s$^{-1}$ are missing. In addition, 
there must be tens of old SNRs in the MCs which are still
to be discovered, since the number of known SNRs in the MCs are smaller than
what is expected \citep{1991ApJ...374..475W}. 
However, the number of discrete sources
necessary to account for the diffuse X-ray emission in the LMC and the SMC
would be $\sim10^{5}$ and $\sim10^{4}$, respectively (see Sect.\,\ref{xlum36}),
magnitudes higher than the expected number of missing sources. Thus 
the contribution of all the faint objects is small.

Furthermore, if
the emission was a superposition of the emission from different sources, we would
expect many thermal components 
which might result in a powerlaw spectrum.
As will be shown in Sect.\,\ref{mcmodel}, the local spectrum of the
diffuse X-ray emission in each analyzed $0\fdg25 \times 0\fdg25$ region
can be well reproduced by one thermal spectrum model. 
Therefore, the diffuse emission can mainly be attributed to hot interstellar gas.

\subsection{Background reduction}\label{backred}

In PSPC observations there is a contamination in lower pulse height channels 
due to afterpulse events
\citep[APE,][]{1993ApJ...418..519P,1994ApJ...424..714S} which are thought
to be caused by negative ion formation close 
to the anodes. 
We eliminated the APE by selecting out all events which occurred within 0.37\,ms 
after and not
further away than few arcmin from a preceding event in channels
lower than 18.

Next, time intervals with higher background were screened out 
according to the housekeeping information of each observation.
One part of the background is caused by X-rays of Solar origin scattering off 
the Earth atmosphere \citep{1993ApJ...404..403S}.
The mainly contributing scattering mechanisms are 
Thomson scattering on oxygen, helium, and molecular nitrogen and fluorescent
scattering of molecular nitrogen (K$\alpha$) and atomic oxygen (K$\alpha$).
The oxygen column density 
\n{O} in the housekeeping data was used as an indicator for the Solar X-ray 
background, giving the oxygen column
density along the line of sight through the Earth's residual atmosphere. 
In order to estimate the values for the low, acceptable 
Solar X-ray background, time intervals 
with Solar zenith angle (angle between the Sun, the Earth, and 
the telescope) larger than 120\degr\ were selected in the housekeeping data. 
This means that the satellite was in the shadow of the Earth and the effect 
of the Solar X-ray scattering was low. Thereafter the 
mean $\langle N_{\rm O} \rangle$ and the standard deviation $\sigma$ of \n{O} 
was computed for these nightside observations. With the help of these numbers,
events in time intervals with 
\n{O} not exceeding the upper limit of  $\langle N_{\rm O} \rangle + 1\sigma$ 
were finally selected from the event files.

Being high energy particle detectors, proportional counters are sensitive
to cosmic rays as well. 
The particles falling through the detector also ionize the
counter gas and cause signals. Therefore, the detector was equipped with an 
additional anode which was linked with the active detector
anode in anti-coincidence, and the vetoing during the observation was measured
\citep[master veto,][]{1992ApJ...393..819S}. 
For preliminary particle background elimination, only intervals with master 
veto rate lower than 170\,cts~s$^{-1}$ were chosen.
Finally, we had to eliminate time intervals with enhanced count rates of 
unknown origin 
(short time and long time enhancements) which can be seen as an
increase of the number of measured photons in the housekeeping data. For this purpose, 
mean values of the count rates (60\,s bin) 
were determined and intervals with
count rate enhancements higher than +1$\sigma$ were excluded.

\subsection{Creating spectra}

\begin{figure}
%\centerline{\fbox{\resizebox{8.5cm}{!}{\includegraphics[clip]{h3553f1a.eps}}}}
\vspace*{5mm}
%\centerline{\fbox{\resizebox{8.5cm}{!}{\includegraphics[clip]{h3553f1b.eps}}}}
\caption{\label{cheesed} Merged PSPC data of the MCs. Because the images 
are not exposure corrected, there are bright regions correlating with the 
pointings. Selected boxes are superimposed. Pointings used for the 
estimation of foreground and background emission are marked as 'BG'.}
\end{figure}

The cleaned event files of the single pointings were merged and binned 
for the SMC and the LMC resulting in a cheesed image for each of
the MCs. 
In total, the LMC data contain 1\,990\,000~cts after the 
screening process, the SMC data 730\,000~cts. The mean exposure time for the
LMC is 8\,ks, the maximum exposure time 66\,ks. 
In the SMC, the mean and maximum exposure times are 10\,ks and 43\,ks,
respectively.  
For a spatially resolved spectral analysis of the diffuse emission from the 
MCs, the data were divided into boxes of the size $0\fdg25 \times 0\fdg25$.
Only boxes containing more than 1000 events were used for the following 
spectral analysis in order to have enough photon statistics. 
In Fig.\,\ref{cheesed} the merged data and the selected regions are shown.

In the reductional steps in Sect.\,\ref{backred}, 
the particle background was eliminated only by
excluding time intervals of high master veto rate. In order to
completely dispose of 
the contribution of the high energy particles, their spectrum was
modeled and subtracted  
from all the analyzed spectra. According to \citet{1992ApJ...393..819S} and
\citet{1993ApJ...418..519P,1996ApJ...458..861P} the particle spectrum consists 
of three components. The first component is the internally 
produced spectrum of particles, measured when the filter wheel of the PSPC is 
closed by a $\sim$2\,mm thick aluminum mask so that no X-rays are penetrating 
($S_{\rm FWC}$). There are two additional components which are measured when
the filter wheel is open, but the detector out of focus: Al K$\alpha$ line at 
$\sim$1.5\,keV which is excited when the filter wheel is open and particles
fall through the optical path of the telescope ($S_{\rm Al}$), and the
externally produced flat increase of the whole spectrum ($S_{\rm Ext}$).
Since they are caused by infalling particles, they are all well correlated
with the master veto rate. 
The components of the particle background spectrum 
as functions of the pulse height channel $CH$ can be parameterized as follows:
\begin{eqnarray}
S_{\rm FWC} & = & \lbrack\,39.66\,{CH}^{-2.91} - 3.96 \times 10^{-6}\,{CH}
 \nonumber \\
& & + 0.0045\,\rbrack\ \times CH^{-1} \label{parback_1}
\\
S_{\rm Al} & = & \lbrack\,0.835\,{CH}^{-0.75} \nonumber \\
& & \times exp{\{ -0.716 (12
.247 - \sqrt{CH})^{2} \}} \rbrack\ \times CH^{-1} \label{parback_2}
\\
S_{\rm Ext} & = & \lbrack\,-4.47 \times 10^{-6}\,{CH} + 0.00493\,\rbrack\ 
 \times CH^{-1}  \label{parback_3}
\end{eqnarray}
Particle background dominates in higher energy ranges, because the cosmic X-ray
background is relatively low and the X-ray mirror reflectivity of the 
telescope decreases significantly in these higher energy ranges. Therefore,
counts 
in pulse height channels higher than 230 can be more or less ascribed to 
high energy particles and are used for normalizing the particle background 
spectrum.
Finally we obtain for the total particle background:
\begin{eqnarray}
S({MV,CH}) & = & \lbrack\,(0.018+7.37 \times 10^{-4}\,{MV}) \times
S_{\rm FWC} \nonumber \\
 & & + (-0.004+2.29 \times 10^{-4}\,{MV}) \times S_{\rm Al} \nonumber 
\\
 & &  + (0.007+2.21 \times 10^{-4}\,{MV}) \times S_{\rm Ext} \rbrack \nonumber\\
 & & \times N_{230} 
 \label{parback_4} 
\end{eqnarray}
\begin{tabbing}
$N_{230}$: \= \kill
$MV$: \> mean master veto rate\\
$N_{230}$: \> normalization of the spectrum with counts in \\
\> channels higher than 230
\end{tabbing}

Events of the PSPC data were binned into spectra in each selected $0\fdg25
\times 0\fdg25$ box. The particle background spectrum was modeled for each
spectrum taking the components $S_{\rm FWC, Al, Ext}$ into account
as defined 
in the Eqs.\ (\ref{parback_1}) to (\ref{parback_3}), and determining the
master veto rate for the selected events. 
After normalizing with the
number of events in the channels above 230, the modeled background spectrum 
(Eq.\,(\ref{parback_4}))
was subtracted from the total spectrum. 

\section{Spectral analysis of the diffuse X-ray emission of the MCs}

\subsection{Absorption of X-rays by the ISM}

X-rays traveling through the ISM are absorbed due to photoionization of 
atomic and molecular gas as well as interstellar dust grains. In the course of 
spectral analysis, the X-ray spectra obtained from observations must be 
corrected for these modifications. 
Latest studies of absorption of X-rays in the ISM are presented
by \citet{2000ApJ...542..914W}. They discuss updated photoionization cross
sections and abundances in the ISM and describe the effect of interstellar
grains in detail. 

For energies higher than $\sim$0.5\,keV, the absorption is dominated by the 
heavier elements.
Thus the X-ray opacity inferred from observations of \ion{H}{i} is a lower 
limit, since substantial additional X-ray absorption is contributed by He, C, 
N, O, etc.\ in the ISM. Therefore the contributions of heavier elements must be
considered, e.g.\ using the Solar metallicity for the Galactic ISM. 
The 
corrections for
H$_{2}$ clouds and grains along the line of
sight are small and can be neglected unless spectral analysis of high accuracy
(1 -- 2\%) is carried out. 

\subsection{Background and foreground components in the spectra}\label{bgfgspec}

Before modeling the spectral contribution of the MCs, components in the 
spectra coming either from the cosmic background or from the Galaxy in the
foreground had to be determined. For this purpose, spectra of regions 
within the
selected 10\degr\ $\times$ 10\degr\ and 5\degr\ $\times$ 5\degr\ fields 
including the LMC and the SMC, respectively, but distant enough from the 
MCs were analyzed first. For the LMC, pointings to the south 
ecliptic pole lying northeast of the LMC were suitable, for the SMC, 
pointings to the
southeast of the SMC 
(see Fig.\,\ref{cheesed}).

For the spectra, the cosmic background in the hard band and thermal emission
from the Galactic foreground in the soft bands were taken into
account. 
The cosmic background is well described by a highly absorbed power-law with 
photon index $\Gamma = 1.4$ as e.g.\ ASCA observations have shown
\citep{1995PASJ...47L...5G,1997MNRAS.285..449C,1998A&A...334L..13M}. 
It is absorbed by heavier elements both in the MCs (see below)
and in the Galaxy 
which varies between \n{H\,Gal} $= 3 \times 10^{20}$ and 
8 $\times 10^{20} {\rm  cm}^{-2}$ depending on position
\citep{1990ARA&A..28..215D}.
 
The foreground is described by two thermal components as a composition 
of the Local Bubble and the Galactic halo emission for which two 
Raymond \& Smith models \citep{1977ApJS...35..419R} were used. 
Solar abundances were always assumed for Galactic components. 
The halo emission is absorbed by the Galactic \n{H\,fg1} = \n{H\,Gal}, 
whereas the absorption of the Local Bubble emission \n{H\,fg2} is 
about one order of magnitude lower. 
From ROSAT observations, the softest component was verified to be thermal  
with a plasma temperature 
between 0.05 and 0.09~keV 
\citep{1998A&A...334L..13M}. In order to obtain
the other parameters of the background+foreground spectrum, the spectra of 
regions as mentioned above were modeled using following components: 
1.\ power-law for the cosmic background (fixed absorbing column densities 
\n{H\,Gal} \& \n{H\,MCtot} and fixed $\Gamma$), 
2.\ Raymond \& Smith spectrum for the Galactic halo (fixed absorption with 
\n{H\,fg1} = \n{H\,Gal}, $k T_{\rm fg1}$ to be determined), and 
3.\ Raymond \& Smith spectrum for the Local Bubble  
(0.05 $\leq k T_{\rm fg2} \leq$ 0.09
and \n{H\,fg2} $\leq$ \n{H\,Gal} for the absorption). 
The fit of the spectra from regions close to the MCs resulted 
in $k T_{\rm fg1} = 0.18$~keV for the halo emission which is in good  
agreement with former ASCA and ROSAT results. 
For the Local Bubble component, the plasma temperature
$k T_{\rm fg2} = 0.09$~keV achieved the best fit. \n{H\,fg2}  
varies between 0.05 and 1.0 $\times 10^{20} {\rm cm}^{-2}$ for the LMC
and between 0.25 and 0.90 $\times 10^{20} {\rm cm}^{-2}$ for the SMC.
\n{H\,fg2} was set variable within these limits later on in spectral
models including background, foreground, and MC emission.
Furthermore, values for the emission measures $EM$ of the spectral components 
were obtained for the LMC and the SMC
individually, and were later used for spectral fitting.

\subsection{Modeling the diffuse emission from the Magellanic Clouds}\label{mcmodel}

\begin{figure}[t]
%\centerline{\resizebox{8.8cm}{!}{\includegraphics[clip]{h3553f2.eps}}}
\caption{\label{sketch} Emission contributing to the 
analyzed spectrum. $S_{\rm MC}$, $S_{\rm BG}$, $S_{\rm halo}$, and 
$S_{\rm LB}$ are non-absorbed original emission components, 
$N_{\rm H}$ is the absorbing
H column density.
}
\end{figure}

All the spectra of the MCs were first fitted with a spectral model only 
containing contributions of the 
background and the foreground in order to verify whether there is additional
emission or not. 
In the next step, the diffuse emission from the
MCs was fitted with an additional absorbed Raymond \& Smith component yielding
the characteristic temperature $T_{\rm MC}$ and the absorption \n{H\,MC}. 
One of the most crucial tasks thereby was to find the best fitting interval
for the temperature value. 
For $k T_{\rm MC} < 0.1$~keV, the foreground absorption of the X-rays becomes  
high, modifying the intrinsic luminosity  by factors of about 10. 
Consequently, additional emission components with small values of $T_{\rm MC}$ 
might be modeled, although in 
reality there is only very low additional emission which should be 
better ascribed to variations of the Galactic foreground emission. 
Therefore, the MC component was restricted to
$k T_{\rm MC} \geq 0.1$~keV. 

As shown in Fig.\,\ref{sketch}, the total observed spectrum is
\begin{equation}\label{totalspec}
S_{\rm obs} = S^{\rm abs}_{\rm MC} + (S^{\rm abs}_{\rm BG} + S^{\rm abs}_{\rm halo} + S^{\rm abs}_{\rm LB}),
\end{equation}
with
\begin{eqnarray}\label{mcspec}
S^{\rm abs}_{\rm MC} & = & e^{-\sigma(E)\,N_{\rm H\,Gal}^{\dagger}(\zeta_{\rm \sun}^{\dagger})} \times e^{-\sigma(E)\,N_{\rm H\,MC}^{\diamondsuit}(\zeta_{\rm MC}^{\dagger})} \nonumber \\
& & \times S_{\rm RS}(T_{\rm MC}^{\diamondsuit},\zeta_{\rm MC}^{\dagger},EM_{\scriptsize \sq}^{\diamondsuit}).
\end{eqnarray}
$S_{\rm RS}$ is the Raymond \& Smith model for thermal emission, 
for $S^{\rm abs}_{\rm BG}, S^{\rm abs}_{\rm halo}, S^{\rm abs}_{\rm LB}$ 
see Eqs.\ (\ref{halospec}), (\ref{lbspec}), (\ref{bgspec}).
In Eqs.\ (\ref{totalspec}) to (\ref{bgspec}) free parameters which 
were to 
be determined are marked
with $\diamondsuit$ and fixed parameters with $\dagger$.
For the abundances in the MCs, correction factors $\zeta$ reported by
\citet{1992ApJ...384..508R} were used: 0.5 for the LMC, and 0.2 for the SMC
(with respect to solar).

\begin{figure}[t]
%\centerline{\resizebox{8.8cm}{!}{\includegraphics[angle=270,clip]{h3553f3.eps}}}
\caption{\label{spectrum} Spectrum of the diffuse X-ray emission of the 
field No 282 in the LMC. The model components for the 
extragalactic background, the Galactic halo, and the Local Bubble 
are plotted as well. 
The dashed line shows the LMC emission component.}
\end{figure}

Again the Galactic foreground was considered as an additional
component using the temperatures $T_{\rm fg1}, T_{\rm fg2}$, emission measures,
and absorption column densities \n{H\,fg1}, \n{H\,fg2} with Solar abundances 
as mentioned above:
\begin{equation}\label{halospec}
S^{\rm abs}_{\rm halo} = e^{-\sigma(E)\,N_{\rm H\,Gal}^{\dagger}(\zeta_{\rm \sun}^{\dagger})} \times S_{\rm RS}(T_{\rm fg1}^{\dagger},\zeta_{\rm \sun}^{\dagger},EM_{\rm fg1}^{\dagger}),
\end{equation}
\begin{equation}\label{lbspec}
S^{\rm abs}_{\rm LB} = e^{-\sigma(E)\,N_{\rm H\,fg2}^{\diamondsuit}(\zeta_{\rm \sun}^{\dagger})} \times S_{\rm RS}(T_{\rm fg2}^{\dagger},\zeta_{\rm \sun}^{\dagger},EM_{\rm fg2}^{\dagger}). 
\end{equation}
The cosmic background was modeled with the values for the photon index 
$\Gamma$ and the emission measure as determined previously, 
but the absorption by both
the MCs (\n{H\,MCtot}) and the Galaxy (\n{H\,fg} = \n{H\,Gal}) was taken
into account:
\begin{eqnarray}\label{bgspec}
S^{\rm abs}_{\rm BG} & = & e^{-\sigma(E)\,N_{\rm H\,Gal}^{\dagger}(\zeta_{\rm \sun}^{\dagger})} \times e^{-\sigma(E)\,N_{\rm H\,MCtot}^{\dagger}(\zeta_{\rm MC}^{\dagger})} \nonumber \\
& & \times S_{\rm pow}(\Gamma^{\dagger},n_{\rm pow}^{\dagger}),
\end{eqnarray}
$S_{\rm pow}$ being the power-law model with normalization $n_{\rm pow}$.
For the total absorbing column density \n{H\,MCtot} of the MCs, 
values from the \ion{H}{i} column density maps were used which were 
derived from the Australia Telescope Compact Array
(ATCA) aperture synthesis mosaic survey and the 64\,m Parkes single dish
telescope observations 
(60\arcsec\ resolution for the LMC: 
Kim et al.\ \citeyear{1998ApJ...503..674K}; 
Kim et al.\ \citeyear{2001...ASP...K}, and 98\arcsec\ 
for the SMC: Stanimirovic et al.\ \citeyear{1999MNRAS.302..417S}). 
In Fig.\,\ref{spectrum} a sample spectrum
is shown with all the model components.

The existence of additional diffuse X-ray emission from the MCs was verified
for each $0\fdg25 \times 0\fdg25$ box, and the model yielding the
lower reduced $\chi^{2}$ was chosen as the correct one. 
For all spectra
with evidences for a MC component, the temperature $T_{\rm MC}$ of the
emitting interstellar plasma, the local absorbing column density in the MCs 
$N_{\rm H\,MC}$, and the emission measure $EM_{\scriptsize \sq}$ 
(see Eq.\,(\ref{xspec_norm}) for definition) for the analyzed boxes were 
obtained for both the LMC and SMC.  
The determined temperatures range from 0.1\,keV up to 1.5\,keV which means
that in both MCs hot gas with temperatures from 10$^{6}$\,K to higher than 
10$^{7}$\,K is found.
The results of the modeling 
are listed in Tables \ref{specpar_lmc} and \ref{specpar_smc}
with box
positions, absorption \n{H\,MC}, temperature $T_{\rm MC}$, emission measure,
and additionally derived physical parameters of the emitting gas.

\begin{figure}[t]
%\centerline{\resizebox{8.5cm}{!}{\includegraphics[]{h3553f4a.eps}}}
\vspace*{3mm}
%\centerline{\resizebox{8.5cm}{!}{\includegraphics[]{h3553f4b.eps}}}
\caption{\label{ktima} 
  Temperature distribution image of the LMC and the SMC. 
  Square regions 
  of the size $0\fdg25 \times 0\fdg25$ selected for spectrum extraction 
  were filled with temperature values determined by spectral fitting. The
  image was then Gauss-filter smoothed. Positions of SNRs observed by ROSAT are
  shown by squares, those of XRBs by crossed squares, and for SSSs
  by double squares.} 
\end{figure}

Since each spectrum 
can be assigned to a certain region in the MC fields, images of the 
fit results were created by filling the square regions with derived
temperature values $T_{\rm MC}$. They are shown in Fig.\,\ref{ktima}. 
In these images the pixel values were set to zero in regions
without data. Therefore, the margin of the regions with $T_{\rm MC} > 0$ is 
mainly caused by the fact that there was no or not enough data for the analysis 
in the adjacent regions (see Fig.\,\ref{cheesed}).  
Also in Figs.\,\ref{lumima} and \ref{dellima},
the regions with pixel values $> 0$ are not
characteristic for the extent of the total diffuse emission, but for the regions
for which data was available.

\subsection{Errors of the spectral parameters}\label{error}

For all spectral parameters of the diffuse X-ray emission from the MCs, 
uncertainties of the background and the foreground 
emission as well as the attenuation of X-rays by matter in the 
foreground cause a 
systematic error in addition to the statistical errors from the spectral fit.
The statistical errors $\sigma_{\rm stat}$ of the parameters 
$T_{\rm MC}$, $EM_{\scriptsize \sq}$, and
$N_{\rm H\,MC}$ were determined by fitting the spectra with all the other 
parameters fixed (background and foreground components, absorption of these 
components, and abundances). For $T_{\rm MC}$ the average statistical error
$\sigma_{\rm stat}$ 
is $-10$\%, $+13$\%, for $EM_{\scriptsize \sq}$ $\sim\pm35$\%, 
and for $N_{\rm H\,MC}$ $\pm57$\%.

\subsubsection{Stray light}\label{stray}

In Fig.\,\ref{ktima} one can see that in the very north of the LMC 
and in the very south,
there are regions with high $T_{\rm MC}$. 
But as can be seen in Fig.\,\ref{lumima} (see Sect.\,\ref{xlum36}),
the significance of the luminosity is relatively low. 
The high temperatures measured in these extended 
regions ($k T_{\rm MC} \ge 1.0$~keV) are caused by the high-mass X-ray
binary LMC X-3 in the north and the low-mass X-ray binary 
LMC X-2 in the south. The stray light from these 
hard X-ray sources could not be eliminated from the analyzed PSPC data, 
making both the temperature and the emission measure higher than what is 
typical for the hot interstellar gas.

\subsubsection{Systematic error}

The estimation of the background and foreground components in the spectrum was 
based on the work on cosmic X-ray background by \citet{1995PASJ...47L...5G}, 
\citet{1997MNRAS.285..449C}, and \citet{1998A&A...334L..13M}. Since ASCA
data was mainly used for the analysis presented by these authors, 
the spectral range which was studied 
extended up to 10~keV. Therefore, reliable information was given especially for
the hard X-ray background: $\Gamma = 1.4 \pm 0.1$. 
For the errors of the foreground components,
the background+foreground emission spectra of regions mentioned in 
Sect.\,\ref{bgfgspec} were analyzed 
by varying the values for $\Gamma$ in the range 
specified by its error. In doing so, the errors of the temperatures of the foreground
components were determined: $k T_{\rm fg1} = 0.18^{+0.05}_{-0.01}$~keV
and $k T_{\rm fg2} = 0.09 \pm 0.01$~keV.

Background and foreground emission components are not the only 
sources of systematic error of spectral modeling, but the uncertainty of the 
strength of X-ray absorption by matter in the line of sight must be considered 
as well. As shown by \citet{2000ApJ...542..914W}, 
the correction for 
contributions by molecules
and grains is negligible. However, if clouds are located in the line of sight, 
the absorption will be higher than estimated from the \ion{H}{i} column 
density maps both for the Galaxy and for the MCs. Since the \ion{H}{i} column 
density maps have moderate spatial resolution 
(1\degr\ for the Galactic foreground), 
single clouds cannot be 
resolved, and the values used for the Galactic foreground absorption 
(\n{H\,fg}) and the absorption of the cosmic background by the MCs 
(\n{H\,MCtot}) are average values.

In order to estimate the systematic error $\sigma_{\rm sys}$ of the spectral 
analysis of the ISM of the MCs, all 
the uncertainties mentioned so far were taken into consideration and  
arbitrarily selected regions were fitted with all possible combinations of the 
lower and upper limits of $\Gamma$, $T_{\rm fg1}$, $T_{\rm fg2}$, \n{H\,fg}, 
and \n{H\,MCtot}. Finally, the systematic errors for the parameters of the 
spectral component assigned to the MC emission were obtained: for the 
temperature $T_{\rm MC}$ the systematic error $\sigma_{\rm sys}$ is 
$\sim-5$\%, $+8$\%, for $EM_{\scriptsize \sq}$ $\sim-10$\%, $+150$\%, 
and for $N_{\rm H\,MC}$ $\sim-5$\%, $+20$\%.

\subsection{Physical properties of the hot ISM}

\subsubsection{X-ray luminosity}\label{xlum36}

\begin{figure}[t]
%\centerline{\resizebox{8.8cm}{!}{\includegraphics[]{h3553f5a.eps}}}
\vspace*{3mm}
%\centerline{\resizebox{8.8cm}{!}{\includegraphics[]{h3553f5b.eps}}}
\caption
{\label{lumima} Images showing the significance of the luminosity of
the diffuse X-ray emission.
The luminosity was derived from the results of the spectral fit
and divided by the 1$\sigma$ error. 
For the LMC the positions of the supergiant shells SGS LMC\,1 to 5 are
marked as ellipses.}
\end{figure}

The unabsorbed luminosity of the gas
in the $0\fdg25 \times 0\fdg25$ box regions of the MCs
can be obtained from the spectral fit based on the 
model of \citet{1977ApJS...35..419R}.
In the same way as the temperature images in 
Fig.\,\ref{ktima} 
were created, images for the luminosity distribution were created by filling 
the square regions with $L_{\scriptsize \sq}$ and smoothing them.  
The luminosity of the hot gas in a $0\fdg25 \times 0\fdg25$ region 
in the ROSAT band (0.1 -- 2.4~keV) ranges from
$8.5 \times 10^{33}$ to $2.6 \times 10^{36}$\,erg\,s$^{-1}$ in the LMC and
from $2.7 \times 10^{32}$ to $9.3 \times 10^{35}$\,erg\,s$^{-1}$ in the SMC.
The total luminosity of the diffuse emission from the MCs was calculated 
by summing up the luminosities $L_{\scriptsize \sq}$ of the 
$0\fdg25 \times 0\fdg25$ regions.
For the LMC, the obtained total luminosity is 
$3.2 \times 10^{38}$\,erg\,s$^{-1}$, for the SMC it is
$1.1 \times 10^{37}$\,erg\,s$^{-1}$, both results
with an uncertainty of $\sigma \approx -40$\%, $+100$\%. 

In Fig.\,\ref{lumima} the significance of the
luminosity determination of the diffuse emission is presented.
The computed values of $L_{\scriptsize \sq}$ are listed in 
Tables \ref{specpar_lmc} and \ref{specpar_smc}. 

\subsubsection{Geometry of the hot gas}\label{geogas}

\begin{figure}[t]
%\centerline{\resizebox{8.5cm}{!}{\includegraphics[]{h3553f6a.eps}}}
\vspace*{3mm}
%\centerline{\resizebox{8.5cm}{!}{\includegraphics[]{h3553f6b.eps}}}
\caption{\label{dellima} The distribution of the depth of the X-ray
  emitting regions in the MCs is shown. Similar to the temperature
  distribution images, after the regions used for spectrum extraction were 
  filled with $\Delta l$, the images were smoothed. Symbols are the same as
  in Fig.\,\ref{ktima}.
}
\end{figure}

After the interstellar gas was heated up by a shock front, the high temperature
is maintained because of the negligible cooling.
Matter is mostly gathered to form a shell, so in its
interior the gas can be treated as a 
perfect 
gas, and the density of the hot
ISM can be estimated using the 
perfect 
gas equation. For coronal gas of cosmic
abundances, the gas consists of 
ions 
and electrons, densities of which are
related as 
$n_{\rm e} = 1.213n$ ($\sim10$\% He plus small fraction of heavier elements).
Since the metalicity is lower than in the 
Solar neighborhood both for the LMC and SMC, 
it is 
$n_{\rm e} = (1.2 + 0.013 \zeta)n$.
Therefore the number density for the gas is 
\begin{equation}\label{dens_ideal}
n = \frac{p}{(2.3 + 0.015 \zeta)~k T}.
\end{equation}
Assuming $p = 3 \times 10^{-18}$~bar
\citep[typical value for the Galactic ISM,][]{1990ARA&A..28...71S}
and $T = 10^{6}$~K, the density is $n = 0.01$~cm$^{-3}$. 

The emission measure  
\begin{equation}\label{xspec_norm}
EM_{\scriptsize \sq} = \frac{1}{10^{14} \times 4\pi D^2} \int n_{\rm e} n {\rm 
  d}V~~[{\rm cm}^{-5}], 
\end{equation}
for the $0\fdg25 \times 0\fdg25$ box was obtained 
from the spectral fit, 
with both the electron density $n_{\rm e}$ and gas density $n$ in [cm$^{-3}$] 
and $D$ being the distance to the source in [cm]. We use $D_{\rm 
LMC} = 50~{\rm kpc} = 1.54 \times 10^{23}~{\rm cm}$ and $D_{\rm 
SMC} = 59~{\rm kpc} = 1.82 \times 10^{23}~{\rm cm}$ 
\citep{1999IAUS..190..569V}.
Since the emission measure depends on the radiating volume,
it can be used to derive the depth of the X-ray emitting region,
while the pressure of the hot gas in the galaxy is assumed to be uniform.
In fact, this is a crude assumption.
If the whole galaxy was in pressure equilibrium, this would result in 
anticorrelation between the ISM temperature and the luminosity. As this
is not observed, there must be pressure variations. 
However, assuming pressure equilibrium 
is a good starting point for the following estimation of the 
emitting gas volume, and will allow us to find deviating regions.

Spectra were extracted for regions of the size $0\fdg25 \times 0\fdg25$. 
Therefore, the emitting volume can be written as
\begin{equation}\label{em_vol}
V_{\scriptsize \sq} = A_{\scriptsize \sq} \Delta l =
\Biggl(\frac{0\fdg25}{~360\degr} 2\pi D\Biggr)^2 \Delta l.
\end{equation}
$\Delta l$ is the thickness of the radiating layer. Based on Eqs.\
(\ref{xspec_norm}), 
(\ref{dens_ideal}), 
and (\ref{em_vol}) the depth $\Delta l$ is
\begin{eqnarray}\label{dens_distrib}
\Delta l &=& \frac{EM_{\scriptsize \sq} \times 10^{14} \times 4\pi D^2}{(1.2 + 
0.013 \zeta)~n^{2}} \Biggl(\frac{720}{\pi D}\Biggr)^{2} \nonumber \\
 &=& \frac{4 \times (2.3 + 0.015 \zeta)^2 \times 720^2 \times 10^{14} \times
  k^2}{(1.2 + 0.013 \zeta)~\pi~p^2}~EM_{\scriptsize \sq}~T_{\rm MC}^2 \nonumber \\
 &=& 2.4 \times 10^{4} \times EM_{\scriptsize \sq}~T_{\rm keV}^2~~[{\rm kpc}] 
\end{eqnarray}
with $p = 3 \times 10^{-18}$~bar and $EM_{\scriptsize \sq}$ in
[cm$^{-5}$]. $T_{\rm keV}$ is the plasma temperature in [keV]. A typical value
of $EM_{\scriptsize \sq} = 0.001$~cm$^{-5}$ and $k T_{\rm keV} = 0.3$~keV yield
$\Delta l = 2.2$~kpc. 
Distributions of $\Delta l$ as derived from the fit values $T_{\rm keV}$ and
$EM_{\scriptsize \sq}$ for $p = 3 \times 10^{-18}$~bar are presented in
Fig.\,\ref{dellima}. The values for $\Delta l$
are also listed in Tables \ref{specpar_lmc} and \ref{specpar_smc}. 

In Fig.\,\ref{dellima} one can see that $\Delta l$ 
is suspiciously high in the east of the LMC, 
including the \ion{H}{ii} region 30 Doradus
as well as the SGSs LMC\,2 and LMC\,3 
(see also Fig.\,\ref{lumima}).
This is due to the wrong assumption that the pressure is the same as in the 
rest of the galaxy. In this very active region with a concentration of SNRs
and stellar associations embedded in dense cold gas (see also 
Fig.\,\ref{hikt_lmc}), the pressure can be assumed to be higher, giving 
$\Delta l \propto p^{-2}$ a more realistic value.

The very high values for $\Delta l$ 
in the very north and in the very south of the LMC
are consequences of the stray light of the bright sources 
LMC X-3 and LMC X-2 (see also Sect.\,\ref{stray}). 

\section{Discussion}

\subsection{Temperature distribution}\label{distrib_hot}

In order to verify the regions with significant diffuse X-ray emission and to
look for correlations between the hot ionized gas and other components within
the galaxies, the temperature distribution images (Fig.\,\ref{ktima}) were 
overlaid on other observations of the MCs, i.e.\ merged
image from PSPC data, DSS image, or \ion{H}{i} column density maps.
Fig.\,\ref{hikt_lmc} shows the \ion{H}{i} column density map of the LMC with
contours of X-ray temperatures, and in Fig.\,\ref{dsskt_smc}, the contours are
plotted on a DSS image of the SMC.

\begin{figure}[t]
%\centerline{\resizebox{8.8cm}{!}{\includegraphics[]{h3553f7.eps}}}
\caption{\label{hikt_lmc} Temperature contours from 0.1\,keV to 0.4\,keV in
steps of 0.1\,keV are superimposed on a \ion{H}{i} map of the LMC 
\citep{1998ApJ...503..674K}
with positions of SNRs observed by ROSAT marked as squares.}
\end{figure}

\begin{figure}[t]
%\centerline{\resizebox{8.8cm}{!}{\includegraphics[]{h3553f8.eps}}}
\caption{
\label{dsskt_smc} Overlay of the temperature distribution as contours from
0.1 to 0.8\,keV in steps of 0.1\,keV on a DSS image of the SMC. Boxes
are SNRs observed by ROSAT and crosses are \ion{H}{ii}
regions.}
\end{figure}

\subsubsection{LMC}

In the LMC, SGSs were found in \ion{H}{i} and H$\alpha$ 
observations with diameters of the order of 1\,kpc, i.e.\ each with a size of 
5\% -- 10\% of the total size of the galaxy. The X-ray results show that
there is an extended hot
region in the eastern part of the LMC with a diameter of about 1\,kpc, 
as can be seen in Fig.\,\ref{ktima}. This region covers the 
SGS LMC\,2 \citep{1980MNRAS.192..365M} 
where many active objects like SNRs and OB associations were found 
(see Figs.\,\ref{ktima}, \ref{lumima}, and \ref{hikt_lmc}). 
Using Einstein data,
\citet{1991ApJ...379..327W} derived a luminosity of $\sim2 
\times 10^{37}$~erg~s$^{-1}$ for the diffuse emission from 
SGS LMC\,2 assuming
a temperature of $\sim5 \times 10^{6}$~K. 
Based on ROSAT PSPC data, \citet{2000ApJ...545..827P} 
get a flux of $1.4 \times 10^{-10}$~erg~cm$^{-2}$~s$^{-1}$ in the energy band
of 0.44 to 2.04~keV, which corresponds to 
$L_{\rm X} = 6 \times 10^{37}$~erg~s$^{-1}$ (0.1 -- 2.4~keV). They used a
model with $k T = 0.31$~keV.
In the same region, we find 
$L_{\rm X} = 1.7 \times 10^{37}$~erg~s$^{-1}$ with a mean temperature of
$k T_{\rm MC} = 0.82$~keV. 

High temperature was also
determined at the northern rim of SGS LMC\,4 and the 
boundary between SGS LMC\,4 and SGS LMC\,5. 
At such boundaries between the hot ISM and the cold dense supergiant shells,
the heating mechanism is most effective, since
the expanding gas hits on dense matter and is strongly decelerated, the
kinetic energy transforming itself to thermal energy. 
For SGS LMC\,4, the integration of the X-ray luminosity gives 
$L_{\rm X} = 1.6 \times 10^{37}$~erg~s$^{-1}$ and the mean value for the 
temperature is $k T_{\rm MC} = 0.24$~keV. This result is  
in a very good agreement with the work by 
\citet{1994A&A...283L..21B}, who obtained a 
temperature of $2.4 \times 10^{6}$~K, i.e.\ $k T = 0.21$~keV for the northern 
part of SGS LMC\,4, as well as with the total luminosity of 
$\sim 2 \times 10^{37}$~erg~s$^{-1}$ for the SGS LMC\,4 which was derived
by \citet{2000ApJ...545..827P} from the results of \citet{1994A&A...283L..21B},
assuming
that the northern emission is representative for the whole SGS LMC\,4.

As for the total diffuse emission of the LMC, \citet{1991ApJ...374..475W}
determined a lower limit of $\sim 2 \times 10^{38}$~erg~s$^{-1}$ for the
luminosity from Einstein data (0.16 -- 3.5~keV). The total luminosity 
derived from the ROSAT PSPC data in the energy band of 0.1 to 2.4~keV
is $3.2 \times 10^{38}$\,erg\,s$^{-1}$ (Sect.\,\ref{xlum36}) with an 
error of $\sim-40$\%, $+100$\%, which is mainly due to uncertainties about
the emission and absorption of the foreground gas.

\subsubsection{SMC}

The hot gas in the SMC is well correlated with the optical main body which 
can be seen on a DSS
image with superimposed temperature contours (Fig.\,\ref{dsskt_smc}). Regions
with the highest temperatures are found in the southwest part of the galaxy
with a large number of SNRs and \ion{H}{ii} regions. The total X-ray emission 
not only arises from the optically visible part of the galaxy,
but also shows the emission from the galactic halo. The existence of diffuse
X-ray emission from the SMC was already
reported by \citet{1991ApJ...377L..85W} based on the analysis of Einstein data 
who obtained a total X-ray luminosity of the diffuse emission 
$L_{\rm X} = 5.0 \times 10^{38}$~erg~s$^{-1}$. But this was not verified in 
later ROSAT observations \citep{1999IAUS..190...32S}. 
As it was shown in Sect.\,\ref{xlum36}, the ROSAT
PSPC data yield  a much lower luminosity 
($L_{\rm X} = 1.1 \times 10^{37}$\,erg\,s$^{-1}$).
A possible explanation for the discrepancy by a factor of about 50
between $L_{\rm X}$ derived from 
Einstein observation and that of ROSAT is the incompleteness of the list  
of the point sources which were removed from the Einstein data, 
since a large number
of new point sources were found in subsequent X-ray observations.

Radio observations
revealed shells and giant shells within the SMC and in the outer regions   
\citep{1997MNRAS.289..225S}, but supergiant shells with large  
cavities like in the LMC are not known. The X-ray emission measured by the 
PSPC is the superposition of the emission from the interior of the 
shells  
which was heated up by stellar winds and supernovae. 

\subsection{Comparison to stellar distribution}

In the LMC, stellar associations forming large scale systems were found
\citep[Shapley's constellations,][]{1953PNAS...39..358M} suggesting secondary
star formation \citep{1983A&A...127..113B,1984A&A...131..347I}. 
The most prominent system is the Shapley's constellation III,
located in the north of the LMC. It coincides spatially with the SGS LMC\,4 and
includes a large number of OB stars in about 20 young associations. The
formation of these stars is thought to be caused by the gravitational
instability of the SGS LMC\,4.
Another well known star formation region in the LMC is the 30 Doradus region
located at the border between SGS LMC\,2 and LMC\,3. 
The comparison of the distribution of young stars in the LMC 
\citep[$t_{\rm age} \leq 2 \times 10^{7}$~yr,][]{1984A&A...131..347I} and the 
diffuse X-ray emission shows that the hot gas is well correlated with the 
distribution of young stars. In regions with highest temperatures, i.e.\ 
around SGS LMC\,4 or in the 30 Doradus region,
there is a concentration of young supergiants 
($t_{\rm age} \leq 8 \times 10^{6}$~yr).

From optical observations it is known, that a young population of stars in 
the SMC is located along the bar \citep[e.g.][]{1980A&A....87...92B}, and in
particular concentrated in the southwest. In this part of the SMC, the hot gas
detected in X-rays coincides with the most active regions. This can be 
verified in Fig.\,\ref{dsskt_smc} showing the positions of \ion{H}{ii} regions,
since associations of massive OB stars form \ion{H}{ii} regions by their ionizing
radiation and winds. Star forming
regions are located in the eastern wing of the SMC as well, extending from the
northern end to the east \citep{1992MNRAS.257..195G}. Hot gas is  
distributed between the wing and the optical main body of the SMC, in a 
region around RA = 01$^{\sl h}$ 04$^{\sl m}$, Dec~=~--72\degr~30\arcmin\ 
surrounded by SNRs and \ion{H}{ii} regions in the north. 
The wing region is thought to be
caused by dynamical interactions between the SMC and the LMC and/or between the
MCs and the Galaxy which triggered star formation 
\citep{1980PASJ...32..581M,1990A&ARv...2...29W}. 
A concentration of gas in the wing region was probably formed
by forces acting between the halos of the MCs, rather than by an outflow of
interstellar material from the main body of the SMC. 

\subsection{Diffuse emission from other nearby galaxies}

\begin{table*}[t]
\caption{\label{othergal}
Properties of diffuse emission from some other nearby galaxies.}
\begin{tabular}{llccccp{5.7cm}l}
\noalign{\smallskip}\hline\noalign{\smallskip}
\multicolumn{1}{c}{1} & \multicolumn{1}{c}{2} & \multicolumn{1}{c}{3} 
& \multicolumn{1}{c}{4} & \multicolumn{1}{c}{5} & \multicolumn{1}{c}{6} 
& \multicolumn{1}{c}{7} \\
\hline\noalign{\smallskip}
\multicolumn{1}{c}{Name} & \multicolumn{1}{c}{Type} 
& \multicolumn{1}{c}{Distance} & \multicolumn{1}{c}{\ion{H}{i} mass} 
& \multicolumn{1}{c}{N$_{\rm H, Gal}$} 
& \multicolumn{1}{c}{$L_{\rm X}$} & \multicolumn{1}{c}{References} \\
 & & \multicolumn{1}{c}{[Mpc]} & \multicolumn{1}{c}{[$10^{09}$~M$_{\sun}$]} 
& \multicolumn{1}{c}{[$10^{20}$~cm$^{-2}$]} 
& \multicolumn{1}{c}{[$10^{39}$~erg~s$^{-1}$]} & \\
\noalign{\smallskip}\hline\noalign{\smallskip}
SMC & SB(s)m & 0.06 & 0.4  & 5.1  & 0.01  & Stanimirovic et al.\ (2000), this work \\
NGC 1705 & SA0 & 5.0 & 0.09  & 3.5  & 0.12  & Hensler et al.\ (1998) \\
LMC & SB(s)m & 0.05 & 0.5  & 6.0  & 0.32 & McGee \& Milton (1966), this work \\
NGC 1569 & IBm & 2.2 & 0.2  & 22  & 0.4  & Heckman et al. (1995) \\
NGC 4449 & IBm & 3.7 & 1.0  &  1.2  & 1.0  & Theis \& Kohle (2001), Vogler \& Pietsch (1997) \\
NGC 4631 & SB(s)d & 7.5 & 7.0  & 1.2  & 4.0  & Rand (1994), Vogler \& Pietsch (1996) \\
NGC 253 & SAB(s)c & 2.6  & 1.0  & 1.3  & 4.0  & Puche et al.\ (1991), Pietsch et al.\ (2000) \\
NGC 4258 & SAB(s)bc & 6.4 & 5.0  & 1.2   & 20  & van Albada \& Shane (1975), Vogler \& Pietsch (1999) \\
M 83 & SAB(s)c & 8.9 & 20  & 4.0  &  36  & Huchtmeier \& Bohnenstengel (1981), Ehle et al.\ (1998) \\
\noalign{\smallskip}
\hline
\end{tabular}
~\\*[2mm]Notes to\\
Col.\ 2:
Obtained from NASA/IPAC Extragalactic Database (NED).\\
Col.\ 5: 
Galactic foreground \nh\ \citep{1990ARA&A..28..215D}. \\
Col.\ 6: 
Luminosity (0.1 -- 2.4~keV) 
of the diffuse emission from the galaxies
corrected for Galactic foreground absorption.\\
\end{table*}

As was shown in the last sections, the diffuse X-ray emission of the MCs 
seems to arise from the halo on the one hand, and from regions with high star
formation activity on the other hand. This is in good agreement with the 
models for stellar bubbles and 
the evolution of stars, predicting that
the ISM is heated up by stellar winds and supernova explosions. 
The hot gas in the ISM can blow out of the galactic disk and flow into the 
halo. Diffuse X-ray emission as evidence for the ISM heating is also 
observed in other nearby galaxies. Some examples are listed in 
Table \ref{othergal}
with distance, \ion{H}{i} mass, foreground \n{H}, and X-ray luminosity.

NGC 1704 and NGC 1569 are Magellanic-type nearby galaxies with ongoing star
formation activities. Their \ion{H}{i} mass is comparable to that of the MCs
\citep{1966AuJPh..19..343M,2000MNRAS.315..791S}.
Similar to the MCs, there are bright star clusters 
embedded in superbubbles, and diffuse X-ray emission shows correlations with 
the H$\alpha$ distribution. The unabsorbed X-ray luminosity of the diffuse 
emission is similar to that of the LMC 
\citep{1998ApJ...502L..17H,1995ApJ...448...98H}.
The face-on irregular galaxy NGC 4449 resembles the LMC as well, regarding 
its size, structure, and the diffuse X-ray emission 
\citep{1997A&A...319..459V,2001A&A...370..365T}.
For the outer disk and the halo, a temperature of $\sim 3 \times 10^{6}$~K 
was measured for the interstellar plasma, created probably during the star 
formation between $2 \times 10^{6}$ and $4 \times 10^{7}$ years ago.

Similarity in diffuse 
X-ray luminosity is also observed between the LMC and spiral
galaxies, like the edge-on spiral galaxy NGC 4631 which is known to have 
many \ion{H}{ii} regions and moderate star formation activity. 
The diffuse X-ray emission \citep{1996A&A...311...35V} 
is about one order of magnitude higher than that of the LMC,
as expected due to the bigger galaxy mass
\citep{1994A&A...285..833R}. Like the MCs, the gas 
distribution in NGC 4631 is thought to have been influenced by tidal 
interactions with its neighboring galaxies \citep{1978A&A....65...47C}. 
In addition, supergiant \ion{H}{i} shells like in the LMC were observed 
\citep{1993AJ....105.2098R}.

One of the most famous nearby galaxies with X-ray emission studied
in detail is the edge-on starburst galaxy NGC 253 with relatively low
\ion{H}{i} mass \citep{1991AJ....101..456P}. \citet{2000A&A...360...24P}
analyzed the ROSAT data and found that the contributions of the nuclear 
area, disk, and halo to the diffuse X-ray emission are about equal. 
The nuclear area mainly consists of a heavily absorbed source with 
$k T = 1.2$~keV and $L_{\rm X} = 3 \times 10^{38}$~erg~s$^{-1}$ and an 
'X-ray plume' described by two components ($k T = 1.2$~keV and 
$k T = 0.33$~keV), which is thought to originate from the interaction between 
the galactic wind from the starburst nucleus and the interstellar medium 
within the disk. The halo emission is very soft with a temperature of 
$k T \approx 0.1$~keV. In the spectra of the MCs,
this component is difficult to be verified because of the higher
Galactic foreground absorption. 

Further spiral galaxies are known to show diffuse X-ray emissions,
like the Seyfert 2 galaxy NGC 4258 
\citep{1975A&A....42..433V,1999A&A...352...64V} or the face-on spiral 
galaxy M 83 \citep{1981A&A...100...72H,1998A&A...329...39E}. 
These galaxies show at least two diffuse emission components, 
the soft halo emission and the hard, highly absorbed disk emission. 
Their diffuse X-ray 
luminosity is very high ($L_{\rm X} > 10^{40}$~erg~s$^{-1}$) what is to be 
expected due to their star formation activity.

\section{Summary and outlook}

Analyzing the ROSAT PSPC data of pointed observations of the MCs,
the temperature distribution of the hot component of the ISM in
the MCs was
determined with a 15\arcmin\ $\times$ 15\arcmin\ resolution.
The hot thin interstellar plasma in the
LMC and the SMC has temperatures of 10$^{6}$ to 10$^{7}$~K. In the LMC, highest
temperatures were determined in regions including the SGSs LMC 2 and LMC 3,
as well as in the northern part of SGS LMC 4. These are the regions with
enhanced star formation rate where young massive stars were found.
In the SMC, the interstellar gas is hottest in the southwestern part.
Many SNRs were detected in this part of the galaxy, which 
is also known to have a large concentration of young stars. 
The results on the SMC is of particular importance, since its
diffuse X-ray emission had not been
studied in detail so far. Furthermore, the X-ray
luminosity profile of the hot ISM was derived. The total
luminosity within the observed area amounts to
$3.2 \times 10^{38}$\,erg\,s$^{-1}$ in the LMC and
$1.1 \times 10^{37}$\,erg\,s$^{-1}$ in the SMC
with $\sigma \approx -40$\%, $+100$\%. 
The X-ray luminosity of the LMC is in the same order as
that of other nearby galaxies which show diffuse emission. These galaxies 
have in 
common that their star formation rate is high. Based on the temperatures
and emission measures as results of spectral modeling, 
the depth of the hot ISM was computed.
The distribution of the hot ISM in the MCs shows that the 
highest temperatures are well correlated with secondary star formation,
whereas the total hot gas is distributed over the whole galaxy.

The ISM is thought to be enriched and heated up during
the stellar evolution, forming high-density shells of cold matter by 
compression. Nevertheless, the physical properties of the thin interstellar 
plasma are still 
not well understood. Thin highly ionized gas with magnetic fields comparable to
those in the 
interstellar space is difficult to create and to be maintained in the 
laboratories on Earth. The new
X-ray telescopes like XMM-Newton or Chandra
with improved sensitivity and spectral resolution make it 
possible to study the thin plasma in interstellar space in full detail and 
determine its physical state.   
Furthermore, observations of other nearby galaxies in the near future
will allow more elaborate studies of single objects and the ISM of galaxies.
The comparison of the results from the analysis of other nearby galaxies to 
those of the MCs will give additional observational proofs for the matter 
recycling within the galaxies and will 
improve the understanding of the evolution of
galaxies.

\begin{acknowledgements}
We would like to thank the anonymous referee for helpful comments.
We are grateful to Drs L.\ Staveley-Smith and S.\ Kim for providing us with 
the \ion{H}{i}
column density maps of the LMC and the SMC.
The ROSAT project is supported by the German
Bundesministerium f\"ur Bildung und Forschung (BMBF) and
the Max-Planck Society.
This research has made use of the NASA/IPAC Extragalactic Database (NED) 
which is operated by the Jet Propulsion Laboratory, California Institute of 
Technology, under contract with the National Aeronautics and Space 
Administration. We acknowledge the use of NASA's {\it SkyView} facility
(http://skyview.gsfc.nasa.gov) located at NASA Goddard Space Flight Center.
This research is 
based on photographic data obtained using The UK Schmidt Telescope.
The UK Schmidt Telescope was operated by the Royal Observatory
Edinburgh, with funding from the UK Science and Engineering Research
Council, until 1988 June, and thereafter by the Anglo-Australian
Observatory.  Original plate material is copyright (c) the Royal
Observatory Edinburgh and the Anglo-Australian Observatory.  The
plates were processed into the present compressed digital form with
their permission.  The Digitized Sky Survey was produced at the Space
Telescope Science Institute under US Government grant NAG W-2166.
\end{acknowledgements}

\clearpage
\textheight23.00cm

\begin{landscape}

\begin{table}
\vspace{1.5cm}
\footnotesize\renewcommand{\arraystretch}{.8} 
\caption{\label{specpar_lmc} 
Spectral fit parameters for the diffuse emission of the LMC.}
\begin{tabular}{rccrccrrcr}
\noalign{\smallskip}\hline\noalign{\smallskip}
\multicolumn{1}{c}{1} & 2 & \multicolumn{1}{c}{3} & \multicolumn{1}{c}{4} & \multicolumn{1}{c}{5} & \multicolumn{1}{c}{6} & \multicolumn{1}{c}{7} & \multicolumn{1}{c}{8} & \multicolumn{1}{c}{9} & \multicolumn{1}{c}{10} \\
\hline\noalign{\smallskip}  
\multicolumn{1}{c}{No} & Center of Region & \multicolumn{1}{c}{No of events} & \multicolumn{1}{c}{$k T_{\rm MC}$} & \multicolumn{1}{c}{$N_{\rm H\,MC}$} & \multicolumn{1}{c}{$EM_{\scriptsize \sq}$} & \multicolumn{1}{c}{red.} & \multicolumn{1}{c}{DOF} & \multicolumn{1}{c}{$L_{\scriptsize \sq}$} & \multicolumn{1}{c}{$\Delta l$} \\
 & \ \ RA\ \ \ \ \ \ \ \ \ \ \ \ Dec & \multicolumn{1}{c}{[10$^{3}$]} & \multicolumn{1}{c}{[keV]} &
 \multicolumn{1}{c}{[10$^{20}$~cm$^{-2}$]} & \multicolumn{1}{c}{[$10^{-3}$~cm$^{-5}$]} & \multicolumn{1}{c}{$\chi^2$} & & \multicolumn{1}{c}{[10$^{36}$~erg/s]} & \multicolumn{1}{c}{[kpc]} \\
 &(J2000.0) &   & & & & & & &   \\
\noalign{\smallskip}\hline\noalign{\smallskip}
   1 & 04 48 21.5~~~--66 04 59 &  1.4 &0.16 (+0.04,--0.02)&  18.31 (+7.55,--10.19)&  2.08 (+0.84,--0.74)& 1.1 &   31 &   0.19  &  1.3 \\ 
   2 & 04 48 42.8~~~--66 35 00 &  2.8 &0.14 (+0.04,--0.02)&  19.87 (+7.65,--11.03)&  1.36 (+0.59,--0.46)& 2.0 &   50 &   0.08  &  0.6 \\ 
   3 & 04 50 37.8~~~--65 50 02 &  1.7 &0.13 (+0.04,--0.02)&  18.64 (+7.57,--10.37)&  2.47 (+0.97,--0.89)& 2.4 &   36 &   0.14  &  1.1 \\ 
   4 & 04 50 49.5~~~--66 05 03 &  2.4 &0.15 (+0.06,--0.02)&   4.14 (+6.54,--2.60)&  0.87 (+0.10,--0.60)& 1.7 &   48 &   0.17  &  0.4 \\
\hline
\end{tabular}
~\\
Notes to Tables \ref{specpar_lmc} and \ref{specpar_smc}:
Results of the spectral analysis of the ROSAT PSPC data with the position of the
regions selected for creating spectrum and the number of events.
For values resulting directly from the spectral fit (plasma temperature $T_{\rm MC}$, absorbing column density of the MCs 
$N_{\rm H\,MC}$, emission measure $EM_{\scriptsize \sq}$) the 90 \% confidence 
range
is given in bracketts. 
\end{table}
\clearpage

\addtocounter{table}{-1}
\begin{table}
\footnotesize\renewcommand{\arraystretch}{.8} 
\caption{Continued (LMC).}
\begin{tabular}{rccrccrrcr}
\noalign{\smallskip}\hline\noalign{\smallskip}
\multicolumn{1}{c}{1} & 2 & \multicolumn{1}{c}{3} & \multicolumn{1}{c}{4} & \multicolumn{1}{c}{5} & \multicolumn{1}{c}{6} & \multicolumn{1}{c}{7} & \multicolumn{1}{c}{8} & \multicolumn{1}{c}{9} & \multicolumn{1}{c}{10} \\
\hline\noalign{\smallskip}  
\multicolumn{1}{c}{No} & Center of Region & \multicolumn{1}{c}{No of events} & \multicolumn{1}{c}{$k T_{\rm MC}$} & \multicolumn{1}{c}{$N_{\rm H\,MC}$} & \multicolumn{1}{c}{$EM_{\scriptsize \sq}$} & \multicolumn{1}{c}{red.} & \multicolumn{1}{c}{DOF} & \multicolumn{1}{c}{$L_{\scriptsize \sq}$} & \multicolumn{1}{c}{$\Delta l$} \\
 & \ \ RA\ \ \ \ \ \ \ \ \ \ \ \ Dec &  \multicolumn{1}{c}{[10$^{3}$]} & \multicolumn{1}{c}{[keV]} &
 \multicolumn{1}{c}{[10$^{20}$~cm$^{-2}$]} & \multicolumn{1}{c}{[$10^{-3}$~cm$^{-5}$]} & \multicolumn{1}{c}{$\chi^2$} & & \multicolumn{1}{c}{[10$^{36}$~erg/s]} & \multicolumn{1}{c}{[kpc]} \\
 &(J2000.0) &   & & & & & & \multicolumn{1}{c}{} & \\
\noalign{\smallskip}\hline\noalign{\smallskip}
  41 & 05 01 36.5~~~--66 50 20 &  4.7 &0.19 (+0.04,--0.02)&   0.00 (+6.25,--0.00)&  1.70 (+0.71,--0.59)& 1.3 &   82 &   0.81  &  1.5 \\ 
  42 & 05 01 55.4~~~--67 05 21 &  3.0 &0.31 (+0.03,--0.03)&   0.00 (+6.25,--0.00)&  0.85 (+0.40,--0.27)& 2.0 &   59 &   0.43  &  2.0 \\ 
  43 & 05 01 56.9~~~--69 05 21 &  1.7 &0.21 (+0.04,--0.02)&   0.00 (+6.25,--0.00)&  1.37 (+0.60,--0.46)& 0.6 &   39 &   0.67  &  1.5 \\ 
  44 & 05 02 15.0~~~--67 20 21 &  2.3 &0.20 (+0.04,--0.02)&   0.00 (+6.25,--0.00)&  1.39 (+0.60,--0.47)& 1.3 &   51 &   0.67  &  1.3 \\
\hline
\end{tabular}
\end{table}

\clearpage

\addtocounter{table}{-1}
\begin{table}
\vspace{2cm}
\footnotesize\renewcommand{\arraystretch}{.8} 
\caption{Continued (LMC).}
\begin{tabular}{rccrccrrcr}
\noalign{\smallskip}\hline\noalign{\smallskip}
\multicolumn{1}{c}{1} & 2 & \multicolumn{1}{c}{3} & \multicolumn{1}{c}{4} & \multicolumn{1}{c}{5} & \multicolumn{1}{c}{6} & \multicolumn{1}{c}{7} & \multicolumn{1}{c}{8} & \multicolumn{1}{c}{9} & \multicolumn{1}{c}{10} \\
\hline\noalign{\smallskip}  
\multicolumn{1}{c}{No} & Center of Region & \multicolumn{1}{c}{No of events} & \multicolumn{1}{c}{$k T_{\rm MC}$} & \multicolumn{1}{c}{$N_{\rm H\,MC}$} & \multicolumn{1}{c}{$EM_{\scriptsize \sq}$} & \multicolumn{1}{c}{red.} & \multicolumn{1}{c}{DOF} & \multicolumn{1}{c}{$L_{\scriptsize \sq}$} & \multicolumn{1}{c}{$\Delta l$} \\
 & \ \ RA\ \ \ \ \ \ \ \ \ \ \ \ Dec &  \multicolumn{1}{c}{[10$^{3}$]} & \multicolumn{1}{c}{[keV]} &
 \multicolumn{1}{c}{[10$^{20}$~cm$^{-2}$]} & \multicolumn{1}{c}{[$10^{-3}$~cm$^{-5}$]} & \multicolumn{1}{c}{$\chi^2$} & & \multicolumn{1}{c}{[10$^{36}$~erg/s]} & \multicolumn{1}{c}{[kpc]} \\
 &(J2000.0) &   & & & & & & \multicolumn{1}{c}{} & \\
\noalign{\smallskip}\hline\noalign{\smallskip}
  81 & 05 09 41.2~~~--71 50 34 &  1.3 &0.17 (+0.04,--0.02)&   0.00 (+6.25,--0.00)&  0.41 (+0.10,--0.29)& 1.3 &   30 &   0.19  &  0.3 \\ 
  82 & 05 09 47.5~~~--70 20 34 &  1.6 &0.22 (+0.02,--0.03)&   0.00 (+6.25,--0.00)&  1.60 (+0.10,--0.24)& 1.3 &   38 &   0.78  &  1.8 \\ 
  83 & 05 09 54.5~~~--68 50 34 &  1.8 &0.22 (+0.03,--0.01)&  19.93 (+7.65,--11.06)&  3.94 (+1.83,--2.67)& 1.9 &   44 &   0.59  &  4.6 \\ 
  84 & 05 10 02.2~~~--67 20 34 &  1.8 &0.15 (+0.01,--0.01)&   0.41 (+4.92,--0.41)&  4.94 (+2.72,--0.72)& 1.5 &   44 &   1.82  &  2.6 \\
\hline
\end{tabular}
\end{table}

\clearpage

\addtocounter{table}{-1}
\begin{table}
\footnotesize\renewcommand{\arraystretch}{.8} 
\caption{Continued (LMC).}
\begin{tabular}{rccrccrrcr}
\noalign{\smallskip}\hline\noalign{\smallskip}
\multicolumn{1}{c}{1} & 2 & \multicolumn{1}{c}{3} & \multicolumn{1}{c}{4} & \multicolumn{1}{c}{5} & \multicolumn{1}{c}{6} & \multicolumn{1}{c}{7} & \multicolumn{1}{c}{8} & \multicolumn{1}{c}{9} & \multicolumn{1}{c}{10} \\
\hline\noalign{\smallskip}  
\multicolumn{1}{c}{No} & Center of Region & \multicolumn{1}{c}{No of events} & \multicolumn{1}{c}{$k T_{\rm MC}$} & \multicolumn{1}{c}{$N_{\rm H\,MC}$} & \multicolumn{1}{c}{$EM_{\scriptsize \sq}$} & \multicolumn{1}{c}{red.} & \multicolumn{1}{c}{DOF} & \multicolumn{1}{c}{$L_{\scriptsize \sq}$} & \multicolumn{1}{c}{$\Delta l$} \\
 & \ \ RA\ \ \ \ \ \ \ \ \ \ \ \ Dec &  \multicolumn{1}{c}{[10$^{3}$]} & \multicolumn{1}{c}{[keV]} &
 \multicolumn{1}{c}{[10$^{20}$~cm$^{-2}$]} & \multicolumn{1}{c}{[$10^{-3}$~cm$^{-5}$]} & \multicolumn{1}{c}{$\chi^2$} & & \multicolumn{1}{c}{[10$^{36}$~erg/s]} & \multicolumn{1}{c}{[kpc]} \\
 &(J2000.0) &   & & & & & & \multicolumn{1}{c}{} & \\
\noalign{\smallskip}\hline\noalign{\smallskip}
 121 & 05 16 10.6~~~--67 50 45 &  2.7 &0.18 (+0.01,--0.01)&   0.00 (+6.25,--0.00)&  3.05 (+0.75,--0.43)& 1.1 &   58 &   1.43  &  2.4 \\ 
 122 & 05 16 17.6~~~--70 35 45 &  1.0 &0.16 (+0.02,--0.04)&   0.00 (+6.25,--0.00)&  2.52 (+0.10,--0.78)& 1.3 &   23 &   1.13  &  1.6 \\ 
 123 & 05 16 23.9~~~--66 35 45 &  1.4 &0.12 (+0.04,--0.02)&   6.74 (+6.73,--3.99)& 10.10 (+2.92,--4.18)& 2.3 &   33 &   1.13  &  3.7 \\ 
 124 & 05 16 28.9~~~--69 20 46 &  2.4 &0.25 (+0.01,--0.01)&   0.00 (+6.25,--0.00)&  4.99 (+0.75,--0.39)& 1.2 &   58 &   2.46  &  7.5 \\
\hline
\end{tabular}
\end{table}

\clearpage

\addtocounter{table}{-1}
\begin{table}
\vspace{2cm}
\footnotesize\renewcommand{\arraystretch}{.8} 
\caption{Continued (LMC).}
\begin{tabular}{rccrccrrcr}
\noalign{\smallskip}\hline\noalign{\smallskip}
\multicolumn{1}{c}{1} & 2 & \multicolumn{1}{c}{3} & \multicolumn{1}{c}{4} & \multicolumn{1}{c}{5} & \multicolumn{1}{c}{6} & \multicolumn{1}{c}{7} & \multicolumn{1}{c}{8} & \multicolumn{1}{c}{9} & \multicolumn{1}{c}{10} \\
\hline\noalign{\smallskip}  
\multicolumn{1}{c}{No} & Center of Region & \multicolumn{1}{c}{No of events} & \multicolumn{1}{c}{$k T_{\rm MC}$} & \multicolumn{1}{c}{$N_{\rm H\,MC}$} & \multicolumn{1}{c}{$EM_{\scriptsize \sq}$} & \multicolumn{1}{c}{red.} & \multicolumn{1}{c}{DOF} & \multicolumn{1}{c}{$L_{\scriptsize \sq}$} & \multicolumn{1}{c}{$\Delta l$} \\
 & \ \ RA\ \ \ \ \ \ \ \ \ \ \ \ Dec &  \multicolumn{1}{c}{[10$^{3}$]} & \multicolumn{1}{c}{[keV]} &
 \multicolumn{1}{c}{[10$^{20}$~cm$^{-2}$]} & \multicolumn{1}{c}{[$10^{-3}$~cm$^{-5}$]} & \multicolumn{1}{c}{$\chi^2$} & & \multicolumn{1}{c}{[10$^{36}$~erg/s]} & \multicolumn{1}{c}{[kpc]} \\
 &(J2000.0) &   & & & & & & \multicolumn{1}{c}{} & \\
\noalign{\smallskip}\hline\noalign{\smallskip}
 161 & 05 20 55.5~~~--66 20 53 &  1.8 &0.26 (+0.03,--0.02)&  12.49 (+7.13,--7.08)&  0.96 (+0.45,--0.31)& 2.1 &   40 &   0.22  &  1.6 \\ 
 162 & 05 20 56.6~~~--67 35 53 &  2.8 &0.23 (+0.02,--0.03)&   0.00 (+6.25,--0.00)&  2.31 (+1.66,--0.19)& 1.7 &   59 &   1.14  &  2.9 \\ 
 163 & 05 20 59.5~~~--68 50 54 &  3.9 &0.30 (+0.03,--0.04)&   3.45 (+8.27,--2.23)&  2.19 (+0.88,--0.78)& 1.4 &   80 &   0.81  &  4.7 \\ 
 164 & 05 21 04.2~~~--70 05 54 &  3.6 &0.25 (+0.02,--0.02)&   0.00 (+6.25,--0.00)&  2.25 (+0.61,--0.21)& 1.3 &   70 &   1.11  &  3.4 \\
\hline
\end{tabular}
\end{table}

\clearpage

\addtocounter{table}{-1}
\begin{table}
\footnotesize\renewcommand{\arraystretch}{.8} 
\caption{Continued (LMC).}
\begin{tabular}{rccrccrrcr}
\noalign{\smallskip}\hline\noalign{\smallskip}
\multicolumn{1}{c}{1} & 2 & \multicolumn{1}{c}{3} & \multicolumn{1}{c}{4} & \multicolumn{1}{c}{5} & \multicolumn{1}{c}{6} & \multicolumn{1}{c}{7} & \multicolumn{1}{c}{8} & \multicolumn{1}{c}{9} & \multicolumn{1}{c}{10} \\
\hline\noalign{\smallskip}  
\multicolumn{1}{c}{No} & Center of Region & \multicolumn{1}{c}{No of events} & \multicolumn{1}{c}{$k T_{\rm MC}$} & \multicolumn{1}{c}{$N_{\rm H\,MC}$} & \multicolumn{1}{c}{$EM_{\scriptsize \sq}$} & \multicolumn{1}{c}{red.} & \multicolumn{1}{c}{DOF} & \multicolumn{1}{c}{$L_{\scriptsize \sq}$} & \multicolumn{1}{c}{$\Delta l$} \\
 & \ \ RA\ \ \ \ \ \ \ \ \ \ \ \ Dec &  \multicolumn{1}{c}{[10$^{3}$]} & \multicolumn{1}{c}{[keV]} &
 \multicolumn{1}{c}{[10$^{20}$~cm$^{-2}$]} & \multicolumn{1}{c}{[$10^{-3}$~cm$^{-5}$]} & \multicolumn{1}{c}{$\chi^2$} & & \multicolumn{1}{c}{[10$^{36}$~erg/s]} & \multicolumn{1}{c}{[kpc]} \\
 &(J2000.0) &   & & & & & & \multicolumn{1}{c}{} & \\
\noalign{\smallskip}\hline\noalign{\smallskip}
 201 & 05 24 59.3~~~--69 21 01 &  8.7 &0.28 (+0.03,--0.02)&   1.57 (+6.36,--1.22)&  3.80 (+1.36,--1.43)& 2.3 &  120 &   1.54  &  7.2 \\ 
 202 & 05 25 01.0~~~--71 36 01 &  2.5 &0.80 (+0.02,--0.04)&  19.95 (+7.65,--11.07)&  0.30 (+0.17,--0.08)& 1.3 &   49 &   0.09  &  4.6 \\ 
 203 & 05 25 02.8~~~--67 06 01 &  1.5 &0.19 (+0.03,--0.04)&   0.00 (+6.25,--0.00)&  1.86 (+6.22,--0.50)& 1.0 &   34 &   0.88  &  1.6 \\ 
 204 & 05 25 18.6~~~--68 21 01 &  5.2 &0.25 (+0.02,--0.02)&  19.61 (+7.63,--10.89)&  3.12 (+0.49,--1.14)& 1.6 &   89 &   0.53  &  4.6 \\
\hline
\end{tabular}
\end{table}

\clearpage

\addtocounter{table}{-1}
\begin{table}
\vspace{2cm}
\footnotesize\renewcommand{\arraystretch}{.8} 
\caption{Continued (LMC).}
\begin{tabular}{rccrccrrcr}
\noalign{\smallskip}\hline\noalign{\smallskip}
\multicolumn{1}{c}{1} & 2 & \multicolumn{1}{c}{3} & \multicolumn{1}{c}{4} & \multicolumn{1}{c}{5} & \multicolumn{1}{c}{6} & \multicolumn{1}{c}{7} & \multicolumn{1}{c}{8} & \multicolumn{1}{c}{9} & \multicolumn{1}{c}{10} \\
\hline\noalign{\smallskip}  
\multicolumn{1}{c}{No} & Center of Region & \multicolumn{1}{c}{No of events} & \multicolumn{1}{c}{$k T_{\rm MC}$} & \multicolumn{1}{c}{$N_{\rm H\,MC}$} & \multicolumn{1}{c}{$EM_{\scriptsize \sq}$} & \multicolumn{1}{c}{red.} & \multicolumn{1}{c}{DOF} & \multicolumn{1}{c}{$L_{\scriptsize \sq}$} & \multicolumn{1}{c}{$\Delta l$} \\
 & \ \ RA\ \ \ \ \ \ \ \ \ \ \ \ Dec &  \multicolumn{1}{c}{[10$^{3}$]} & \multicolumn{1}{c}{[keV]} &
 \multicolumn{1}{c}{[10$^{20}$~cm$^{-2}$]} & \multicolumn{1}{c}{[$10^{-3}$~cm$^{-5}$]} & \multicolumn{1}{c}{$\chi^2$} & & \multicolumn{1}{c}{[10$^{36}$~erg/s]} & \multicolumn{1}{c}{[kpc]} \\
 &(J2000.0) &   & & & & & & \multicolumn{1}{c}{} & \\
\noalign{\smallskip}\hline\noalign{\smallskip}
 241 & 05 28 59.3~~~--66 36 08 &  7.7 &0.24 (+0.02,--0.04)&   9.61 (+6.93,--5.53)&  1.77 (+0.74,--0.61)& 1.8 &  105 &   0.42  &  2.4 \\ 
 242 & 05 29 03.7~~~--70 51 08 &  1.1 &0.23 (+0.04,--0.06)&   0.00 (+6.25,--0.00)&  1.35 (+0.10,--0.34)& 1.0 &   25 &   0.67  &  1.8 \\ 
 243 & 05 29 09.4~~~--65 36 08 &  5.1 &0.19 (+0.04,--0.02)&   3.05 (+6.47,--2.02)&  2.27 (+0.90,--0.81)& 2.5 &   85 &   0.68  &  2.0 \\ 
 244 & 05 29 11.1~~~--69 51 09 &  6.8 &0.30 (+0.01,--0.01)&   0.00 (+3.36,--0.00)&  4.55 (+0.63,--0.25)& 1.4 &  113 &   2.29  &  9.6 \\
\hline
\end{tabular}
\end{table}

\clearpage

\addtocounter{table}{-1}
\begin{table}
\footnotesize\renewcommand{\arraystretch}{.8} 
\caption{Continued (LMC).}
\begin{tabular}{rccrccrrcr}
\noalign{\smallskip}\hline\noalign{\smallskip}
\multicolumn{1}{c}{1} & 2 & \multicolumn{1}{c}{3} & \multicolumn{1}{c}{4} & \multicolumn{1}{c}{5} & \multicolumn{1}{c}{6} & \multicolumn{1}{c}{7} & \multicolumn{1}{c}{8} & \multicolumn{1}{c}{9} & \multicolumn{1}{c}{10} \\
\hline\noalign{\smallskip}  
\multicolumn{1}{c}{No} & Center of Region & \multicolumn{1}{c}{No of events} & \multicolumn{1}{c}{$k T_{\rm MC}$} & \multicolumn{1}{c}{$N_{\rm H\,MC}$} & \multicolumn{1}{c}{$EM_{\scriptsize \sq}$} & \multicolumn{1}{c}{red.} & \multicolumn{1}{c}{DOF} & \multicolumn{1}{c}{$L_{\scriptsize \sq}$} & \multicolumn{1}{c}{$\Delta l$} \\
 & \ \ RA\ \ \ \ \ \ \ \ \ \ \ \ Dec &  \multicolumn{1}{c}{[10$^{3}$]} & \multicolumn{1}{c}{[keV]} &
 \multicolumn{1}{c}{[10$^{20}$~cm$^{-2}$]} & \multicolumn{1}{c}{[$10^{-3}$~cm$^{-5}$]} & \multicolumn{1}{c}{$\chi^2$} & & \multicolumn{1}{c}{[10$^{36}$~erg/s]} & \multicolumn{1}{c}{[kpc]} \\
 &(J2000.0) &   & & & & & & \multicolumn{1}{c}{} & \\
\noalign{\smallskip}\hline\noalign{\smallskip}
 281 & 05 33 24.0~~~--65 21 16 &  5.4 &0.15 (+0.01,--0.01)&  19.99 (+7.66,--5.47)&  6.40 (+1.55,--1.99)& 1.9 &   81 &   0.46  &  3.5 \\ 
 282 & 05 33 24.3~~~--67 21 16 &  2.7 &0.25 (+0.02,--0.03)&   0.00 (+8.53,--0.00)&  2.62 (+0.10,--0.30)& 1.6 &   61 &   1.29  &  3.8 \\ 
 283 & 05 33 25.4~~~--64 21 16 &  1.1 &0.29 (+0.03,--0.02)&   0.00 (+6.25,--0.00)&  0.30 (+0.17,--0.08)& 1.5 &   24 &   0.15  &  0.6 \\ 
 284 & 05 33 26.6~~~--68 21 16 &  3.7 &0.29 (+0.06,--0.01)&  13.11 (+7.16,--7.41)&  1.53 (+0.65,--0.52)& 1.6 &   71 &   0.38  &  3.1 \\
\hline
\end{tabular}
\end{table}

\clearpage

\addtocounter{table}{-1}
\begin{table}
\vspace{2cm}
\footnotesize\renewcommand{\arraystretch}{.8} 
\caption{Continued (LMC).}
\begin{tabular}{rccrccrrcr}
\noalign{\smallskip}\hline\noalign{\smallskip}
\multicolumn{1}{c}{1} & 2 & \multicolumn{1}{c}{3} & \multicolumn{1}{c}{4} & \multicolumn{1}{c}{5} & \multicolumn{1}{c}{6} & \multicolumn{1}{c}{7} & \multicolumn{1}{c}{8} & \multicolumn{1}{c}{9} & \multicolumn{1}{c}{10} \\
\hline\noalign{\smallskip}  
\multicolumn{1}{c}{No} & Center of Region & \multicolumn{1}{c}{No of events} & \multicolumn{1}{c}{$k T_{\rm MC}$} & \multicolumn{1}{c}{$N_{\rm H\,MC}$} & \multicolumn{1}{c}{$EM_{\scriptsize \sq}$} & \multicolumn{1}{c}{red.} & \multicolumn{1}{c}{DOF} & \multicolumn{1}{c}{$L_{\scriptsize \sq}$} & \multicolumn{1}{c}{$\Delta l$} \\
 & \ \ RA\ \ \ \ \ \ \ \ \ \ \ \ Dec &  \multicolumn{1}{c}{[10$^{3}$]} & \multicolumn{1}{c}{[keV]} &
 \multicolumn{1}{c}{[10$^{20}$~cm$^{-2}$]} & \multicolumn{1}{c}{[$10^{-3}$~cm$^{-5}$]} & \multicolumn{1}{c}{$\chi^2$} & & \multicolumn{1}{c}{[10$^{36}$~erg/s]} & \multicolumn{1}{c}{[kpc]} \\
 &(J2000.0) &   & & & & & & \multicolumn{1}{c}{} & \\
\noalign{\smallskip}\hline\noalign{\smallskip}
 321 & 05 36 52.7~~~--68 36 23 &  2.8 &0.27 (+0.03,--0.02)&   0.68 (+6.30,--0.68)&  2.78 (+1.06,--1.02)& 2.0 &   69 &   1.22  &  4.9 \\ 
 322 & 05 37 03.3~~~--65 51 23 &  5.4 &0.23 (+0.04,--0.02)&   6.35 (+6.70,--3.79)&  1.61 (+0.68,--0.55)& 2.4 &   87 &   0.44  &  2.0 \\ 
 323 & 05 37 06.0~~~--69 36 24 & 27.7 &0.94 (+0.02,--0.05)&  19.99 (+7.66,--11.09)&  4.42 (+1.54,--1.69)& 2.1 &  171 &   1.14  & 93.9 \\ 
 324 & 05 37 12.9~~~--66 51 24 &  2.7 &0.29 (+0.03,--0.03)&   0.70 (+6.30,--0.70)&  0.75 (+0.36,--0.23)& 1.2 &   55 &   0.33  &  1.5 \\
\hline
\end{tabular}
\end{table}

\clearpage

\addtocounter{table}{-1}
\begin{table}
\footnotesize\renewcommand{\arraystretch}{.8} 
\caption{Continued (LMC).}
\begin{tabular}{rccrccrrcr}
\noalign{\smallskip}\hline\noalign{\smallskip}
\multicolumn{1}{c}{1} & 2 & \multicolumn{1}{c}{3} & \multicolumn{1}{c}{4} & \multicolumn{1}{c}{5} & \multicolumn{1}{c}{6} & \multicolumn{1}{c}{7} & \multicolumn{1}{c}{8} & \multicolumn{1}{c}{9} & \multicolumn{1}{c}{10} \\
\hline\noalign{\smallskip}  
\multicolumn{1}{c}{No} & Center of Region & \multicolumn{1}{c}{No of events} & \multicolumn{1}{c}{$k T_{\rm MC}$} & \multicolumn{1}{c}{$N_{\rm H\,MC}$} & \multicolumn{1}{c}{$EM_{\scriptsize \sq}$} & \multicolumn{1}{c}{red.} & \multicolumn{1}{c}{DOF} & \multicolumn{1}{c}{$L_{\scriptsize \sq}$} & \multicolumn{1}{c}{$\Delta l$} \\
 & \ \ RA\ \ \ \ \ \ \ \ \ \ \ \ Dec &  \multicolumn{1}{c}{[10$^{3}$]} & \multicolumn{1}{c}{[keV]} &
 \multicolumn{1}{c}{[10$^{20}$~cm$^{-2}$]} & \multicolumn{1}{c}{[$10^{-3}$~cm$^{-5}$]} & \multicolumn{1}{c}{$\chi^2$} & & \multicolumn{1}{c}{[10$^{36}$~erg/s]} & \multicolumn{1}{c}{[kpc]} \\
 &(J2000.0) &   & & & & & & \multicolumn{1}{c}{} & \\
\noalign{\smallskip}\hline\noalign{\smallskip}
 361 & 05 41 11.4~~~--69 06 31 & 15.1 &0.84 (+0.02,--0.02)&  19.99 (+7.66,--3.73)&  3.42 (+0.12,--0.12)& 1.0 &  149 &   1.01  & 57.8 \\ 
 362 & 05 41 11.9~~~--67 21 31 &  1.2 &0.23 (+0.05,--0.04)&   4.70 (+6.58,--2.90)&  1.73 (+0.10,--0.54)& 1.9 &   27 &   0.53  &  2.3 \\ 
 363 & 05 41 15.5~~~--70 51 32 &  3.7 &0.18 (+0.04,--0.02)&  19.89 (+7.65,--11.04)&  7.15 (+1.38,--1.21)& 1.5 &   75 &   0.80  &  5.9 \\ 
 364 & 05 41 23.0~~~--63 51 31 &  3.8 &1.00 (+0.02,--0.05)&  19.98 (+7.65,--11.09)&  0.94 (+0.44,--0.30)& 2.0 &   71 &   0.23  & 22.7 \\
\hline
\end{tabular}
\end{table}

\clearpage 

\begin{table}
\vspace{2cm}
\footnotesize\renewcommand{\arraystretch}{.8} 
\caption{\label{specpar_smc}
Spectral fit parameters for the diffuse emission of the SMC.}
\begin{tabular}{rccrccrrcr}
\noalign{\smallskip}\hline\noalign{\smallskip}
\multicolumn{1}{c}{1} & 2 & \multicolumn{1}{c}{3} & \multicolumn{1}{c}{4} & \multicolumn{1}{c}{5} & \multicolumn{1}{c}{6} & \multicolumn{1}{c}{7} & \multicolumn{1}{c}{8} & \multicolumn{1}{c}{9} & \multicolumn{1}{c}{10} \\
\hline\noalign{\smallskip}  
\multicolumn{1}{c}{No} & Center of Region & \multicolumn{1}{c}{No of events} & \multicolumn{1}{c}{$k T_{\rm MC}$} & \multicolumn{1}{c}{$N_{\rm H\,MC}$} & \multicolumn{1}{c}{$EM_{\scriptsize \sq}$} & \multicolumn{1}{c}{red.} & \multicolumn{1}{c}{DOF} & \multicolumn{1}{c}{$L_{\scriptsize \sq}$} & \multicolumn{1}{c}{$\Delta l$} \\
 & \ \ RA\ \ \ \ \ \ \ \ \ \ \ \ Dec & \multicolumn{1}{c}{[10$^{3}$]} & \multicolumn{1}{c}{[keV]} &
 \multicolumn{1}{c}{[10$^{20}$~cm$^{-2}$]} & \multicolumn{1}{c}{[$10^{-3}$~cm$^{-5}$]} & \multicolumn{1}{c}{$\chi^2$} & & \multicolumn{1}{c}{[10$^{36}$~erg/s]} & \multicolumn{1}{c}{[kpc]} \\
 &(J2000.0) & & & & & & & & \\
\noalign{\smallskip}\hline\noalign{\smallskip}
   1 & 00 36 39.8~~~--73 14 19 &  2.3 &0.14 (+0.03,--0.01)&   2.22 (+7.22,--1.11)&  1.37 (+10.41,--0.19)& 1.4 &   37 &   0.24  &  0.6\\ 
   2 & 00 40 33.2~~~--73 44 26 &  2.3 &0.12 (+0.03,--0.01)&   5.58 (+10.58,--2.79)&  0.62 (+10.19,--0.09)& 1.3 &   37 &   0.06  &  0.2\\ 
   3 & 00 41 59.8~~~--75 14 29 &  5.9 &0.19 (+0.03,--0.02)&   9.97 (+10.97,--5.99)&  2.41 (+10.72,--0.34)& 2.0 &   68 &   0.07  &  2.2\\ 
   4 & 00 44 07.5~~~--73 44 32 &  2.4 &0.11 (+0.03,--0.01)&   9.96 (+10.96,--5.98)&  0.01 (+10.00,--0.00)& 1.6 &   36 &   0.00  &  0.0\\
\hline
\end{tabular}
\end{table}

\clearpage

\addtocounter{table}{-1}
\begin{table}
\vspace{2cm}
\footnotesize\renewcommand{\arraystretch}{.8} 
\caption{Continued (SMC).}
\begin{tabular}{rccrccrrcr}
\noalign{\smallskip}\hline\noalign{\smallskip}
\multicolumn{1}{c}{1} & 2 & \multicolumn{1}{c}{3} & \multicolumn{1}{c}{4} & \multicolumn{1}{c}{5} & \multicolumn{1}{c}{6} & \multicolumn{1}{c}{7} & \multicolumn{1}{c}{8} & \multicolumn{1}{c}{9} & \multicolumn{1}{c}{10}  \\
\hline\noalign{\smallskip}  
\multicolumn{1}{c}{No} & Center of Region & \multicolumn{1}{c}{No of events} & \multicolumn{1}{c}{$k T_{\rm MC}$} & \multicolumn{1}{c}{$N_{\rm H\,MC}$} & \multicolumn{1}{c}{$EM_{\scriptsize \sq}$} & \multicolumn{1}{c}{red.} & \multicolumn{1}{c}{DOF} & \multicolumn{1}{c}{$L_{\scriptsize \sq}$} & \multicolumn{1}{c}{$\Delta l$} \\
 & \ \ RA\ \ \ \ \ \ \ \ \ \ \ \ Dec & \multicolumn{1}{c}{[10$^{3}$]} & \multicolumn{1}{c}{[keV]} &
 \multicolumn{1}{c}{[10$^{20}$~cm$^{-2}$]} & \multicolumn{1}{c}{[$10^{-3}$~cm$^{-5}$]} & \multicolumn{1}{c}{$\chi^2$} & & \multicolumn{1}{c}{[10$^{36}$~erg/s]} & \multicolumn{1}{c}{[kpc]} \\
 &(J2000.0) & & & & & & & & \\
\noalign{\smallskip}\hline\noalign{\smallskip}
  41 & 01 04 02.5~~~--71 30 09 &  3.0 &0.24 (+0.03,--0.02)&   9.52 (+10.52,--5.26)&  0.27 (+10.08,--0.04)& 1.4 &   40 &   0.01  &  0.4\\ 
  42 & 01 04 24.9~~~--73 15 10 &  6.7 &0.19 (+0.03,--0.02)&   6.14 (+7.14,--3.07)&  0.49 (+10.15,--0.07)& 1.3 &   63 &   0.02  &  0.4\\ 
  43 & 01 04 32.6~~~--71 45 10 &  7.3 &0.21 (+0.03,--0.02)&   9.05 (+10.05,--5.52)&  1.93 (+10.58,--0.27)& 1.7 &   68 &   0.07  &  2.1\\ 
  44 & 01 05 03.7~~~--72 00 11 &  3.5 &0.32 (+0.03,--0.03)&   9.86 (+10.86,--5.93)&  3.53 (+11.06,--0.49)& 1.3 &   59 &   0.25  &  8.7\\
\hline
\end{tabular}
\end{table}

\end{landscape}

\end{document}